\documentclass[sigconf]{acmart}

\usepackage{graphicx}
\usepackage{textcomp}
\usepackage{adjustbox}
\usepackage{xcolor}
\usepackage{booktabs}
\usepackage{multirow}
\usepackage[inkscapelatex=false]{svg}

\AtBeginDocument{%
  }

\copyrightyear{2025}
\acmYear{2025}
\setcopyright{acmlicensed}
\acmConference[WWW Companion '25]{Companion Proceedings of the ACM Web Conference 2025}{April 28-May 2, 2025}{Sydney, NSW, Australia}
\acmBooktitle{Companion Proceedings of the ACM Web Conference 2025 (WWW Companion '25), April 28-May 2, 2025, Sydney, NSW, Australia}
\acmDOI{10.1145/3701716.3717566}
\acmISBN{979-8-4007-1331-6/2025/04}

\begin{document}

\title{Image Fusion for Cross-Domain Sequential Recommendation}

\author{Wangyu Wu}
\affiliation{%
  \institution{Xi’an Jiaotong-Liverpool University}
  \institution{The University of Liverpool}
  \city{Suzhou}
  \country{China}}
\email{v11dryad@foxmail.com}

\author{Siqi Song}
\affiliation{%
  \institution{Xi’an Jiaotong-Liverpool University}
  \institution{The University of Liverpool}
  \city{Suzhou}
  \country{China}}
\email{Siqi.Song22@student.xjtlu.edu.cn}

\author{Xianglin Qiu}
\affiliation{%
  \institution{Xi’an Jiaotong-Liverpool University}
  \institution{The University of Liverpool}
  \city{Suzhou}
  \country{China}}
\email{Xianglin.Qiu20@student.xjtlu.edu.cn}

\author{Xiaowei Huang}
\affiliation{%
  \institution{The University of Liverpool}
  \city{Liverpool}
  \country{UK}}
\email{xiaowei.huang@liverpool.ac.uk}

\author{Fei Ma}
\authornote{Corresponding Author}
\affiliation{%
  \institution{Xi’an Jiaotong-Liverpool University}
    \city{Suzhou}
  \country{China}}
\email{fei.ma@xjtlu.edu.cn}

\author{Jimin Xiao}
\authornotemark[1]
\affiliation{%
  \institution{Xi’an Jiaotong-Liverpool University}
    \city{Suzhou}
  \country{China}}
\email{jimin.xiao@xjtlu.edu.cn}

\renewcommand{\shortauthors}{Wangyu Wu et al.}

\begin{abstract}
Cross-Domain Sequential Recommendation (CDSR) aims to predict future user interactions based on historical interactions across multiple domains. The key challenge in CDSR is effectively capturing cross-domain user preferences by fully leveraging both intra-sequence and inter-sequence item interactions. In this paper, we propose a novel method, Image Fusion for Cross-Domain Sequential Recommendation (IFCDSR), which incorporates item image information to better capture visual preferences. Our approach integrates a frozen CLIP model to generate image embeddings, enriching original item embeddings with visual data from both intra-sequence and inter-sequence interactions. Additionally, we employ a multiple attention layer to capture cross-domain interests, enabling joint learning of single-domain and cross-domain user preferences. To validate the effectiveness of IFCDSR, we re-partitioned four e-commerce datasets and conducted extensive experiments. Results demonstrate that IFCDSR significantly outperforms existing methods.
\end{abstract}




\maketitle

\section{Introduction}
\label{sec:intro}

Sequential Recommendation (SR) has gained significant attention as a method to model dynamic user preferences. SR~\cite{yang2023debiased} aims to recommend the next item of interest for users by analyzing their historical interaction sequences. However, single-domain data can lead to biased recommendations. To address this issue and provide a more comprehensive understanding of user behavior, Cross-Domain Sequential Recommendation (CDSR)~\cite{pinet,yuan2024dual,dagcn,yuan2024balancing,recguru,xu2023multi} has been introduced. CDSR leverages information from multiple domains to improve recommendation accuracy.
\begin{figure}[t]
\centering
\includegraphics[width=0.9\linewidth]{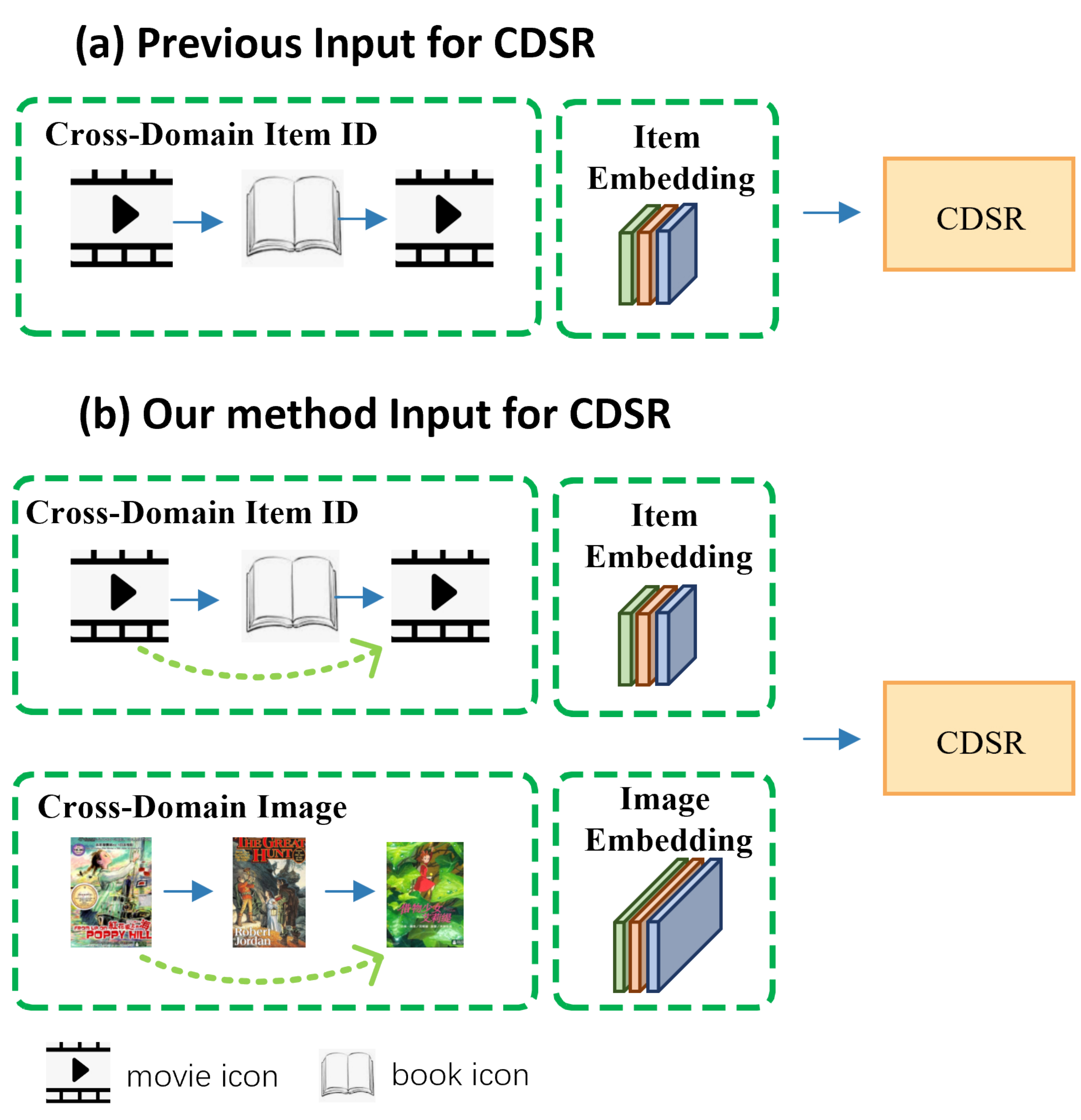}
\vspace{-0.3cm} 
\caption{(a) In the traditional CDSR framework, the input consists solely of item ID features. (b) In our IFCDSR framework, we incorporate additional image information to complement the existing item features.}
\label{fig:idea}
\vspace{-0.5cm} 
\end{figure}
Pioneering work in Cross-Domain Sequential Recommendation (CDSR), such as $\pi$-Net~\cite{pinet}, models item interaction sequences within a single domain to generate representations, which are then transferred to other domains via a gated transfer module. MIFN~\cite{mifn} further incorporates an external knowledge graph module to strengthen the connections across domains, providing a richer context for cross-domain interactions. However, existing works~\cite{pinet, PSJnet} primarily focus on modeling item relationships within sequences from a single domain, often neglecting the inter-domain relationships that are crucial for effective cross-domain recommendations. As a result, these models tend to suffer from a significant form of domain bias—wherein the model's assumptions about item interactions are heavily influenced by the characteristics of a single domain. This issue is particularly prevalent in Sequential Recommendation (SR) tasks, where domain bias is often limited to specific user-item interactions within a particular domain. In contrast, CDSR domain bias stems from the lack of proper alignment between domains, leading to the misrepresentation of cross-domain interactions and suboptimal performance when transferring knowledge across domains.

Additionally, existing CDSR methods do not effectively incorporate the visual representations of the items. We observe that users often form their first impressions of an item through its visual representation. Even when items share similar titles or item features, their visual representations can vary significantly. This variability can greatly influence user interest. To address these limitations, we propose a method called Image Fusion for Cross-Domain Sequential Recommendation (IFCDRS, see Fig.~\ref{fig:idea}). By integrating image embeddings that align with user preferences, we can complement existing item features, enriching item representations. Additionally, IFCDRS learns single-domain and cross-domain user preferences in a joint way, further enhancing recommendation performance. \textbf{Our main contributions are summarized as follows: }

\begin{figure*}[t] 
\begin{center}
   \includegraphics[width=1.0\linewidth]{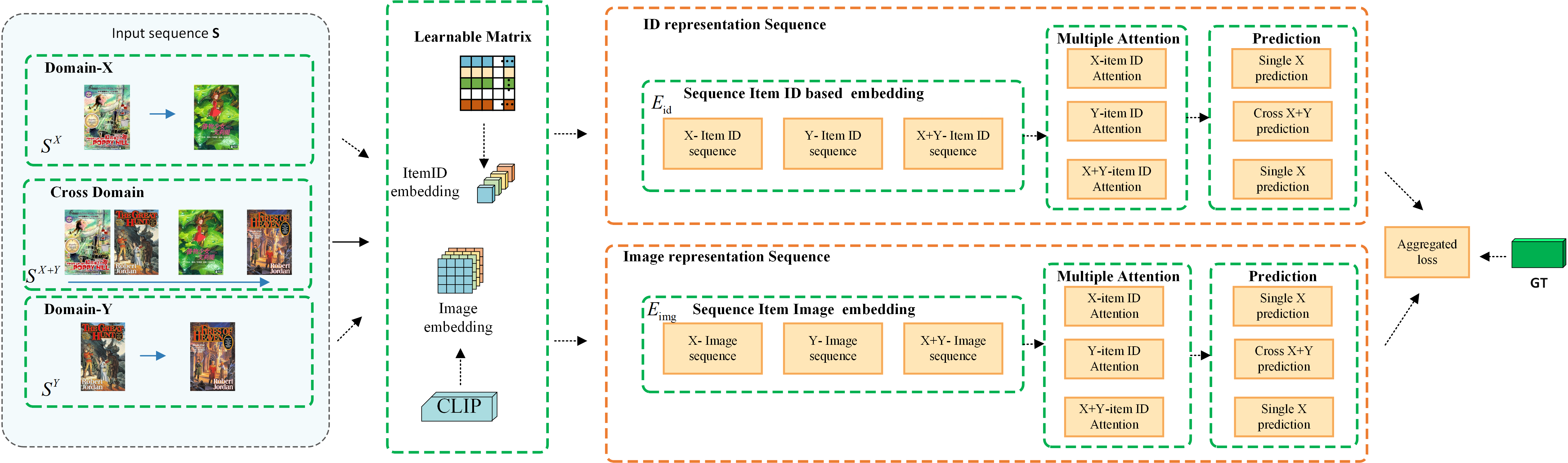}

   \caption{The overview of our proposed IFCDSR. Items from domains $X$ and $Y$ are embedded using a learnable ID-based matrix $E_{id}$ and a frozen CLIP image encoder $E_{img}$. The input sequence $\mathcal{S}$, comprising $S^X$, $S^Y$, and $S^{X+Y}$, is embedded for item ID and image features, then processed through multiple attention layers to capture intra- and inter-sequence relationships. Attention-aggregated embeddings are used with cosine similarity against $E_{id}$ and $E_{img}$ to predict the next item.}
    \label{fig:framework}
\end{center}
\vspace{-0.3cm}
\end{figure*}

\begin{itemize}
    \item We are the first to integrate image features with learnable item embeddings in CDSR, enriching the item representations and enhancing overall recommendation performance.
    \item We introduce a multi-attention method that combines intra-sequence and inter-sequence relationships with image and item features, improving next-item prediction accuracy.
    \item We conduct extensive experiments on two CDSR datasets, and the results demonstrate that our IFCDRS method significantly outperforms existing state-of-the-art baseline methods across all metrics.
\end{itemize}

\section{Related Work}

\subsection{Sequential Recommendation}
Sequential Recommendation~\cite{SRs} serves as the core of our approach, where a user's past behaviors are represented as a sequence of time-sensitive items, with the goal of predicting the likelihood of the next item in the sequence. Early models, such as FPMC~\cite{rendle2010factorizing}, leverage the Markov chain to capture sequential patterns in item interactions. As the field progressed, researchers adopted advanced deep learning techniques, including recurrent neural networks~\cite{GRU, LSTM}, convolutional neural networks~\cite{CNN}, and attention mechanisms~\cite{vaswani2017attention,guo2024dual-hybrid}, to model item dependencies more effectively in recommender systems~\cite{gru4rec, kang2018self, DIN}. However, RNN- and CNN-based approaches typically prioritize recent items when predicting the next item, which limits their ability to capture long-term user preferences. To address this, hybrid models that combine sequential recommendation techniques with traditional models, such as matrix factorization~\cite{koren2009matrix}, have been proposed to balance short-term and long-term interests~\cite{SLIREC, zhao2018plastic}. Additionally, models like SURGE~\cite{SURGE} use metric learning to reduce the dimensionality of item sequences. More recently, methods like DFAR~\cite{lin2023dual} and DCN~\cite{lin2022dual} have emerged, focusing on learning complex relationships within sequential recommendation tasks. In this work, we extend sequential recommendation models to enable cross-domain learning, facilitating knowledge transfer across different domains.

\subsection{Cross-Domain Recommendation}
Cross-domain recommender systems~\cite{CDR} offer a promising approach to address challenges such as data sparsity and the cold-start problem, which are commonly encountered in sequential recommendation tasks. Early models in cross-domain recommendation were built upon single-domain recommendation frameworks, assuming that incorporating auxiliary user behavior data from different domains could enhance user modeling in the target domain. These early approaches primarily relied on transfer learning~\cite{Transfer}, which allows for the transfer of user or item embeddings from the source domain to the target domain, improving model performance. Notable examples include MiNet~\cite{MiNet}, CoNet~\cite{CoNet}, and itemCST~\cite{itemCST}, among others. 

Despite these advances, industrial platforms often aim to improve multiple domains simultaneously, rather than focusing solely on enhancing the target domain without considering the source domain. As a result, dual learning~\cite{jiang2022edsf,long_dual_2012, he_dual_2016,jiang2023structure}, which enables mutual improvement in both source and target domains, has garnered significant attention. This approach has been successfully applied to cross-domain recommender systems~\cite{DTCDR, DDTCDR}. Additionally, to further boost recommendation performance across domains, some researchers have proposed dual-target methods that focus on sequential modeling~\cite{pinet,chen2021dual}. These approaches effectively address the sparse data and cold-start issues while considering the performance of both source and target domains simultaneously. 

For instance, PiNet~\cite{pinet} addresses the shared account problem by transferring account information from one domain to another where the account also has historical behavior. DASL~\cite{DASL} introduces dual embeddings for interaction and dual attention mechanisms to combine sequential patterns across domains for the same user. Similarly, DAT-MDI~\cite{chen2021dual} applies dual attention in session-based recommendations, but without relying on user overlap between domains. However, the assumption that item sequences from two domains can be paired as input is problematic, as item sequences from distinct domains may be independent, even when associated with the same user. The theoretical performance of such dual attention approaches, which mix sequence embeddings from two domains, may not be promising in non-overlapping user scenarios. In contrast, while NATR~\cite{NATR} attempts to avoid user overlap, it remains a non-sequential and single-target model.

\subsection{CLIP-Based Approaches}

The integration of self-supervised learning with vision-language pre-training has significantly advanced the development of CLIP-based models, which aim to bridge the gap between visual task~\cite{wu2024image,li2024high-fidelity,wu2024top} and textual task~\cite{cha2023learning, guo2025underwater,yin2024uncertainty,wu2025adaptive}. These models build upon the original CLIP architecture, incorporating various innovations to enhance representation quality and improve the alignment between images and their corresponding text. One example is SLIP~\cite{mu2022slip}, which improves upon CLIP by incorporating self-supervised learning with image-to-image contrastive learning, producing more robust and comprehensive visual representations. MaskCLIP~\cite{dong2023maskclip} further extends this idea by integrating masked image modeling, which refines visual features by selectively focusing on certain regions of the image. This allows for a more precise alignment of visual features with the associated textual data. A-CLIP~\cite{yang2023attentive} goes a step further by implementing an attention-based token removal strategy, which selectively preserves tokens that are semantically relevant to the text, thus improving the accuracy of the visual-textual alignment.

Although these methods enhance the original CLIP framework by improving the efficiency and quality of representation learning, they primarily focus on global alignment between images and text. This emphasis on global features can limit their effectiveness for tasks that require attention to fine-grained visual details~\cite{yin2023semi, wei2023iclip}. In many of these models, the removal of tokens is often done randomly or based solely on textual cues~\cite{dong2023maskclip,guo2022assessing}. In contrast, DetailCLIP addresses these challenges by employing an attention-based mechanism that takes into account both textual information and fine-grained visual details, leading to improved performance in tasks where preserving such details is crucial.

\section{Methodology}
\label{sec:method}

\textbf{Problem Formulation}: 
In the CDSR task, user interaction sequences occur in domain $X$ and domain $Y$, with their item sets denoted as $\mathcal{X}$ and $\mathcal{Y}$, respectively. Let $\mathcal{S}$ represent the overall interaction sequence of the user in chronological order, which consists of three sub-sequences ${(S^X, S^Y, S^{X+Y}) \in \mathcal{S}}$. Specifically, $S^X = {\left [ x_1, x_2,\dots , x_ {\left | S^X \right |}  \right ]}, {x\in \mathcal{X}}$ and $S^Y = {\left [ y_1, y_2,\dots , y_ {\left | S^Y \right |}  \right ]}, {y\in \mathcal{Y}}$ are the interaction sequences within each domain, where ${\left | \cdot  \right | }$ denotes the total item number. Additionally, ${S^{X+Y} = S^X \cup S^Y}$ represents the merged sequence by combining $S^X$ and $S^Y$. In general, the goal of the CDSR task is to predict the probabilities of  candidate items across both domains and select the item with the highest probability as the next  recommendation. 

\textbf{Overall framework}: As shown in Fig~\ref{fig:framework}, the overview of our IFCDSR framework is depicted. In the beginning, we have an Item Presentations Preparation process that takes all items from domain $X$ and domain $Y$ as input. Each of the items is composed of both item IDs and images. A learnable item matrix $E_{id}$ is initialized to embed all items based on their IDs, while the frozen CLIP image encoder generates the image matrix $E_{img}$ by embedding all items based on their images. 

Subsequently, $\mathcal{S}$ is fed into the network for the Prediction Process. $\mathcal{S}$ consists of three sub-sequences $S^X$, $S^Y$, and $S^{X+Y}$. An embedding layer is applied to generate item ID-based and image-based sequence embeddings. It utilizes $E_{id}$ and $E_{img}$, along with the item sequences in $S^X$, $S^Y$, and $S^{X+Y}$. The corresponding item embeddings are retrieved and organized according to the sequence order, reformulating the respective sequence embeddings. These sequence embeddings are then passed through the multiple attention layer to capture both intra- and inter-sequence relationships, producing attention-aggregated sequence embeddings for each sub-sequence. Finally, these attention-aggregated sequence embeddings are compared with $E_{id}$ and $E_{img}$ using cosine similarity to predict the next item.

\subsection{Image Feature Integration}
\label{sec:Image}
We enhance user image interests by integrating CLIP image features into the CDSR framework. This subsection provides a detailed explanation of feature preparation and sequence representation combined with image features within a single domain, and applies this approach to domain $X$, domain $Y$, and domain $X+Y$.

\subsubsection{Item Presentations Preparation with Image Feature} First, we prepare the feature representations for all items. As shown on the left in Fig.~\ref{fig:framework}, we construct a learnable item matrix based on item IDs for all items in domain $X$ and $Y$, denoted as $E_{id} \in \mathbb{R}^{(|\mathcal{X}|+|\mathcal{Y}|) \times q}$. Here, $|\mathcal{X}|+|\mathcal{Y}|$ represents the total number of items from domains $X$ and $Y$, and $q$ is the learnable item embedding dimension. Simultaneously, we employ a pre-trained and frozen CLIP model to generate image embeddings for each item. These embeddings are used to form image matrix $E_{img} \in \mathbb{R}^{(|\mathcal{X}|+|\mathcal{Y}|) \times e}$, where $e$ is the image embedding dimension. These two matrices are the base for generating embedding for sequences.

\subsubsection{Enhanced Similarity Score with Image Feature}
We first generate a sequence representation to represent the user interaction sequence, which is then used to calculate the similarity with items for predicting the next item in the Prediction Process. As shown on the right in Fig.~\ref{fig:framework}, we propose an embedding layer to process the input sequences $\mathcal{S}$, which consists of three sub-sequences $S^X$, $S^Y$ and $S^{X+Y}$. These sequences include both item IDs and images data. Accordingly, the embedding layer produces both item ID-based and image-based embeddings for each input sequence. Once we get $E_{id}$ and $E_{img}$, the item ID-based sequence embeddings $F_{id} \in \mathbb{R}^{|\mathcal{S}| \times q}$ for $\mathcal{S}$ can be generated by placing appropriate embedding from $E_{id}$ in the order of the items in the sequence $\mathcal{S}$. Here, $|\mathcal{S}|$ denotes the total item number in $\mathcal{S}$. Similarly, the image-based sequence embeddings $F_{img} \in \mathbb{R}^{|\mathcal{S}| \times e}$ are produced from $E_{img}$ based on the same sequence. For simplicity, we take the image-based sequence embeddings $F_{img}$ as an example to illustrate how IFCDSR obtains the next item prediction. We employ an Attention Layer to get the enhanced sequence image-based embeddings $H_{img} \in \mathbb{R}^{|\mathcal{S}| \times e}$ as follows:

\begin{equation}
    H_{img} = Attention(F_{img}),
\end{equation}
Afterward, we extract the last embedding vector $h_{img} \in \mathbb{R}^{1 \times e}$ from $H_{img}$ as the sequence representation, which serves as the attention-aggregated embedding of the sequence. This vector captures the user preferences, specifically focusing on the most recent interaction within the sequence. Then, $h_{img}$ is compared against $E_{img}$ using cosine similarity. In this way, the alignment between user preferences and the embedding of each item across all domains is assessed. The similarity score is computed as follows:

\begin{equation} 
\begin{aligned}\label{eq:cos}
Sim(h_{img}, E_{img}) = \frac{h_{img} \cdot E_{img}^{T}}{\|h_{img}\|\|E_{img}^T\|},
\end{aligned}
\end{equation}
where $T$ denotes the transpose function. Here, a higher similarity score indicates that the sequence preference is more aligned with this item. In the same manner, we can obtain the item ID-based sequence representative vector $h_{id} \in \mathbb{R}^{1 \times q}$ and sequence similarity score. In the next section, we will introduce how the multiple attention layer is used to fuse both ID-based and image-based predictions for CDSR.

\subsection{Multiple Attention Mechanisms}
\label{sec:Attention}

In previous methods~\cite{gru4rec,sasrec}, sequences combining behaviors from domain \( X \) and domain \( Y \) were used without distinction. This can lead to dominance by one domain with a higher volume of behaviors, potentially skewing the results. To address this issue, we design multiple attention mechanisms for the sequence embedding of \( S^X \), \( S^Y \), and \( S^{X+Y} \) to obtain the prediction probabilities for the next item in each sequence as described in Section~\ref{sec:Image}, ensuring balanced representation. Additionally, we combine the ID and image sequences to further enhance the modeling of user interests. The final user representation can be expressed through a set of six sequence representations derived from multiple attention layers: \( h^{X}_{id} \), \( h^{X}_{img} \), \( h^{Y}_{id} \), \( h^{Y}_{img} \), \( h^{X+Y}_{id} \), and \( h^{X+Y}_{img} \). Taking domain \( X \) as an example, the prediction probabilities for the next item in sequence \( S^X \), based on item ID-based sequence embedding, can be expressed as follows:
\begin{equation}
\small
\mathrm{P}^X_{id}(x_{t+1} \mid \mathcal{S}) = softmax(Sim(h_{id}^X, E_{id}^X))
 {\ } {\ } x_t\in {S^X},
\end{equation}
where $x_{t+1}$ denotes the predicted next item, which belongs to the set $\mathcal{X}$ of domain $X$. $h_{id}^X$ and $E_{id}^X$ represent the item ID-based sequence representative vector and item matrix computed within $S^{X}$, respectively. Similarly, the prediction probabilities for the next item based on image embeddings are calculated as:
\begin{equation}
\small
\mathrm{P}^X_{img}(x_{t+1} \mid \mathcal{S}) = softmax(Sim(h_{img}^X, E_{img}^X))
{\ } x_t\in{S^X},
\end{equation}
where $h_{img}^X$ and $E_{img}^X$ denote the item image-based sequence representative vector and image matrix computed within $S^{X}$. Therefore, the final prediction probabilities for the next item $P^X(x_{t+1} \mid \mathcal{S})$ are obtained by summing these two types of probabilities:
\begin{equation}
\small
\mathrm{P}^X(x_{t+1} \mid \mathcal{S}) = \alpha{P}^X_{id}(x_{t+1} \mid \mathcal{S}) + (1-\alpha){P}^X_{img}(x_{t+1} \mid \mathcal{S}),
\end{equation}
where $\alpha$ is a weighting factor between 0 and 1. This approach captures both the structural and visual aspects of items in domain $X$. We employ a standard training approach to optimize the whole framework as follows:

\begin{equation}
\small
	\begin{split}
		\mathcal{L}^X &= \sum_{x_t \in {S^X}}-\log \mathrm{P}^X(x_{t+1} \mid \mathcal{S}).
	\end{split}
	\label{sec_math_eq:single}
\end{equation}
For the remaining sub-sequences \( S_{Y} \) and \( S_{X+Y} \), the same method can be applied to obtain the predictive probability for the next item. The final loss is given by:
\begin{equation}
\mathcal{L} = \mathcal{L}^{X} + \lambda_{1}\mathcal{L}^{Y} + \lambda_{2}\mathcal{L}^{X+Y},
\end{equation}
where $\lambda_1$, and $\lambda_2$ balance the contributions of different losses.

During the evaluation stage, we combine the probabilities from domain \( X \), domain \( Y \), and their joint domain \( X+Y \) to predict the next item. Here, \( X \) is set as the target domain and \( Y \) as the source domain:
\begin{equation}
\small
\mathrm{P}(x_i|S) = \mathcal{P}^{X}(x_i|S) + \lambda_{1} \mathcal{P}^{Y}(x_i|S) + \lambda_{2} \mathcal{P}^{X+Y}(x_i|S).
\end{equation}
Next, we select the item with the highest prediction score in the target domain \( X \) as the recommended next item:
\begin{equation}
\mathrm{argmax}_{x_i \in \mathcal{X}} \mathrm{P}(x_i|S).
\end{equation}


\section{Experiments}
\label{sec:Experiments}

In this section, we introduce the experimental setup, providing detailed descriptions of the dataset, evaluation metrics, and implementation specifics. We then compare our method with state-of-the-art approaches on the Amazon dataset~\cite{wei2021contrastive}. Finally, we conduct ablation studies to validate the effectiveness of the proposed method.

\textbf{Dataset and Evaluated Metric.} Our experiments are conducted on the Amazon dataset~\cite{wei2021contrastive} to construct the CDSR scenarios. Following previous works~\cite{pinet, mifn}, we select four domains to generate two CDSR scenarios for our experiments: ``Food-Kitchen'' and ``Movie-Book''. We first extract users who have interactions in both domains. Then we filter out users and items with fewer than 10 interactions. Additionally, to meet the sequential constraints, we retain cross-domain interaction sequences that contain at least three items from each domain. In the training/validation/test split, the latest interaction sequences are equally divided into the validation and test sets, while the remaining interaction sequences are used for the training set. The statistics of our refined datasets for the CDSR scenarios are summarized in Table \ref{sec_exp_tab:dataset}. We adopt Mean Reciprocal Rank (MRR)~\cite{mrr} and Normalized Discounted Cumulative Gain (NDCG@)\{5, 10\}~\cite{ndcg} as evaluation metrics to assess the performance of our model. These metrics are widely used in recommendation systems.

\begin{table}[!h]
\centering
\caption{Statistics of two CDSR scenarios.}
\setlength{\tabcolsep}{5pt}
\resizebox{8.5cm}{!}{
\begin{tabular}{cccccc}
\toprule
\textbf{Scenarios}  &\textbf{\#Items} &\textbf{\#Train} &\textbf{\#Valid}  &\textbf{\#Test} &\textbf{Avg.length}\\ \midrule
Food  &29,207  &\multirow{2}{*}{34,117} &2,722   &2,747   &\multirow{2}{*}{9.91}     \\ 
Kitchen  &34,886 &\multirow{2}{*}{}  &5,451   &5,659  &\multirow{2}{*}{}     \\
\midrule
Movie  &36,845  &\multirow{2}{*}{58,515} &2,032  &1,978  &\multirow{2}{*}{11.98}    \\ 
Book  &63,937 &\multirow{2}{*}{}  &5,612  &5,730 &\multirow{2}{*}{}   \\ 
\midrule

\end{tabular}
}
\label{sec_exp_tab:dataset}
\end{table}

\textbf{Implementation Details.} For fair comparisons, we adopt the same hyperparameter settings as in previous works~\cite{mifn}. We set the learnable embedding size $q$ to 256, the CLIP-based image embedding size $e$ to 512, $\alpha$ to 0.7, $\lambda_{1}$ to 0.1, and $\lambda_{2}$ to 0.4. The mini-batch size is 256, with a dropout rate of 0.3. The $L_2$ regularization coefficient is selected from $\{0.0001, 0.00005, 0.00001\}$, and the learning rate is chosen from $\{0.001, 0.0005, 0.0001\}$. Training is conducted for 100 epochs using an NVIDIA 4090 GPU, with the Adam optimizer~\cite{adam} to update all parameters.

\begin{table}[h!]
\footnotesize
\centering
\caption{Experimental results (\%) on the Food-Kitchen scenario.}
\label{tab:foodkitchen}
\setlength\tabcolsep{4.5pt} 
\resizebox{\columnwidth}{!}{ 
\begin{tabular}{lcccccc}
\toprule
\multirow{2}{*}{Model (Food-Kitchen)} &
\multirow{2}{*}{MRR} &\multicolumn{2}{c}{NDCG} & \multirow{2}{*}{MRR} &\multicolumn{2}{c}{NDCG} \\
\cmidrule(r){3-4}\cmidrule(l){6-7} &(Food)& @5 & @10  &(Kitchen)& @5  & @10   \\
\midrule
GRU4Rec~\cite{gru4rec}   &   5.79  &   5.48   &  6.13 &
 3.06 &  2.55  &  3.10  \\
SASRec~\cite{sasrec}   &   7.30  &   6.90   &  7.79  &
 3.79 &  3.35  &  3.93  \\ 
SR-GNN~\cite{srgnn}   &    7.84  &   7.58  & 8.35   &
  4.01 &   3.47   &  4.13  \\

MIFN~\cite{mifn}  & 8.55  &   8.28  & 9.01   &
  4.09 &   3.57   &  4.29  \\
Tri-CDR~\cite{ma2024triple} &  8.35 &  8.18 &   8.89  &  
4.29  &  3.63 &   4.33  \\
\textbf{Ours IFCDSR} &\textbf{9.05} &\textbf{8.85} &\textbf{9.92}
&\textbf{4.95} &\textbf{4.56} &\textbf{5.24}\\
\bottomrule
\end{tabular}
}
\end{table}

\begin{table}[h!]
\footnotesize
\centering
\caption{Experimental results (\%) on the Movie-Book scenario.}
\label{tab:moviebook}
\setlength\tabcolsep{4.5pt} 
\resizebox{\columnwidth}{!}{ 
\begin{tabular}{lcccccc}
\toprule
\multirow{2}{*}{Model (Movie-Book)} &
\multirow{2}{*}{MRR} &\multicolumn{2}{c}{NDCG} & \multirow{2}{*}{MRR} &\multicolumn{2}{c}{NDCG} \\
\cmidrule(r){3-4}\cmidrule(l){6-7} &(Movie)  & @5 & @10  &(Book)& @5  & @10   \\
\midrule

GRU4Rec~\cite{gru4rec}   &  3.83 &   3.14 &  3.73  &
 1.68 &  1.34   &  1.52  \\

SASRec~\cite{sasrec}   &   3.79 &   3.23 &  3.69  &
 1.81 &  1.41   &  1.71  \\

SR-GNN~\cite{srgnn}   &  3.85 &  3.27   &  3.78 &
  1.78 &  1.40   &  1.66  \\

PSJNet~\cite{PSJnet}  &  4.63 &  4.06 &   4.76 & 
 2.44  &  2.07 &   2.35  \\

MIFN~\cite{mifn}  &  5.05 &  4.21 &   5.20  &  
2.51  &  2.12 &   2.31  \\
Tri-CDR~\cite{ma2024triple} &  5.15 &  4.62 &   5.05  &  
2.32  &  2.08 &   2.22  \\
\textbf{Ours IFCDSR} &\textbf{6.08}  &\textbf{5.02} &\textbf{5.86}  
&\textbf{2.75} &\textbf{2.37} &\textbf{2.65} \\
\bottomrule
\end{tabular}
}
\end{table}
\subsection{Performance Comparisons}

We compare our result with existing state-of-the-art methods on the ``Food-Kitchen" and `Movie-Book” CDSR scenarios in Tab.~\ref{tab:foodkitchen} and Tab.~\ref{tab:moviebook}, respectively. Our method outperforms existing SOTA approaches in the final performance.

\begin{table}[h!]
\centering
\caption{Ablation study on the movie dataset}
\label{tab:ablation}
\begin{adjustbox}{width=1.0\linewidth}
\begin{tabular}{ccccccc}
\toprule
 original-framework & image fusion & multiple attention & MRR \\
\midrule
\checkmark & &  & 5.03 \\
 \checkmark & \checkmark & &  5.63 \\
 \checkmark & \checkmark & \checkmark & 6.08 \\
\bottomrule
\end{tabular}
\end{adjustbox}
\end{table}

\subsection{Ablation Studies}

We conducted an ablation study to assess the effects of our two key contributions: image fusion and multiple attention mechanisms. As shown in Tab.~\ref{tab:ablation}, we first integrate image fusion into the framework without applying multiple attention blocks for sub-sequences. In this setting, all sequences were combined into a single representation, resulting in a 0.6\% performance improvement. Next, we introduced multiple attention mechanisms to better capture the distinct features of each domain and signal, leading to a further 0.45\% improvement. Overall, our proposed IFCDSR method significantly enhances CDSR performance.

\section{Conclusion}

In this work, we introduce a novel method, Image Fusion for Cross-Domain Sequential Recommendation (IFCDSR). Unlike existing approaches that rely solely on the temporal features of user interaction sequences, we integrate image representations by leveraging the powerful visual understanding capabilities of the CLIP model. Frozen image embeddings are generated and combined with a learnable item matrix to enhance item representations. Additionally, we design a multiple attention layer to capture user sequential information from different domains and sources (image-based and ID-based), further improving the overall performance. As a result, our IFCDSR method achieves state-of-the-art performance in CDSR.
\section{ACKNOWLEDGMENT}
This work was supported by the National Natural Science Foundation of China (No. 62471405, 62331003, 62301451), Suzhou Basic Research Program (SYG202316), XJTLU REF-22-01-010, and the SIP AI Innovation Platform (YZCXPT2022103).

\bibliographystyle{ACM-Reference-Format}
\bibliography{refs}

\end{document}